\documentclass{elsart}

\usepackage{graphicx}
\usepackage{color}

\usepackage{amssymb}
\newcommand{\ac}[1]{c:\Sigma\times\Sigma^{\leq{#1}}\rightarrow\Delta^{+}}
\newcommand{\eac}{\overline{c}:\Sigma^{*}\rightarrow\Delta^{*}}
\newcommand{\sstring}[1]{\sigma_{1}\sigma_{2}\ldots\sigma_{#1}}

\definecolor{myred}{rgb}{1,0.3,0.3}

\begin{document}
\begin{frontmatter}


\title{Translating the EAH Data Compression Algorithm into Automata Theory
	\thanksref{CNCSIS}}


\author{Drago\c s Trinc\u a}
\address{Faculty of Computer Science, ``Al.I.Cuza'' University, 700483 Iasi, Romania}
\author{}\vspace*{-1.55\baselineskip}
\ead{dragost@infoiasi.ro}
\thanks[CNCSIS]{The research reported in this paper was partially supported by CNCSIS grant 632/2004.}

\begin{abstract}
	Adaptive codes have been introduced in \cite{t1} as a new class of non-standard variable-length codes.
	These codes associate variable-length codewords to symbols being encoded depending on the
	previous symbols in the input data string.
	A new data compression algorithm, called EAH,
	has been introduced in \cite{t3},
	where we have behaviorally
	shown that for a large class of input data strings, this algorithm substantially
	outperforms the well-known Lempel-Ziv universal data compression algorithm. In this paper, we translate
	the EAH encoder into automata theory.
\end{abstract}

\end{frontmatter}

\section{Introduction}
	New algorithms for data compression, based on adaptive codes of order one \cite{t1} and Huffman codes,
	have been introduced in \cite{t2}, where we have behaviorally shown that for a large class of input
	data strings, these algorithms substantially outperform the well-known Lempel-Ziv universal data
	compression algorithm \cite{zl1}.

	EAH (Encoder based on Adaptive codes and Huffman codes) has been introduced in \cite{t3}, as an
	improved version of the algorithms presented in \cite{t2}. The work carried out so far \cite{t2,t3}
	has behaviorally proved that EAH is a highly promising data compression algorithm, as one can remark
	again in the examples presented in the following sections. In this paper, we translate the EAH algorithm
	into automata theory.

	Before ending this introductory section, let us recall some basic notions and notations used
	throughout the paper.
	We denote by $|S|$ the \textit{cardinality} of the set $S$; if $x$ is a string of
	finite length, then $|x|$ denotes the length of $x$.
	The \textit{empty string} is denoted by $\lambda$.

	For an alphabet $\Delta$,
	we denote by $\Delta^{*}$ the set
	$\bigcup_{n=0}^{\infty}\Delta^{n}$,
	and by $\Delta^{+}$ the set
	$\bigcup_{n=1}^{\infty}\Delta^{n}$,
	where $\Delta^{0}$ denotes the set $\{\lambda\}$.
	Also, we denote by
	$\Delta^{\leq n}$ the set
	$\bigcup_{i=0}^{n}\Delta^{i}$,
	and by $\Delta^{\geq n}$ the set
	$\bigcup_{i=n}^{\infty}\Delta^{i}$.

	Let $X$ be a finite and nonempty subset of $\Delta^{+}$, and $w\in\Delta^{+}$.
	A \textit{decomposition of w} over $X$ is any sequence of words
	$u_{1}, u_{2}, \ldots, u_{h}$ with $u_{i}\in X$, $1\leq i\leq h$, 
	such that $w=u_{1}u_{2}\ldots u_{h}$.
	A \textit{code} over $\Delta$ is any nonempty set $C\subseteq\Delta^{+}$ such
	that each word $w\in\Delta^{+}$ has at most one decomposition over $C$.
	A \textit{prefix code} over $\Delta$ is any code $C$ over $\Delta$ such that
	no word in $C$ is proper prefix of another word in $C$.
	If $u,v$ are two strings, then we denote by $u\cdot{v}$, or simply by $uv$ the catenation between
	$u$ and $v$.

\section{A Short Review of Adaptive Codes}
	Adaptive codes have been recently introduced in \cite {t1}
	as a new class of non-standard variable-length codes.
	The aim of this section is to briefly review some basic
	definitions, results, and notations directly related to this class of codes.
\begin{defn}
	Let $\Sigma$ and $\Delta$ be alphabets. A function
	$\ac{n}$, $n\geq{1}$, is called \textup{adaptive code of order $n$} if its unique
	homomorphic extension $\eac$ defined by:
\begin{itemize}
\item $\overline{c}(\lambda)=\lambda$ 
\item $\overline{c}(\sstring{m})=$
	$c(\sigma_{1},\lambda)$
	$c(\sigma_{2},\sigma_{1})$
	$\ldots$
	$c(\sigma_{n-1},\sstring{n-2})$
	\newline
	$c(\sigma_{n},\sstring{n-1})$
	$c(\sigma_{n+1},\sstring{n})$
	$c(\sigma_{n+2},\sigma_{2}\sigma_{3}\ldots\sigma_{n+1})$
	\newline
	$c(\sigma_{n+3},\sigma_{3}\sigma_{4}\ldots\sigma_{n+2})\ldots$
	$c(\sigma_{m},\sigma_{m-n}\sigma_{m-n+1}\ldots\sigma_{m-1})$
\end{itemize}
	for all $\sstring{m}\in\Sigma^{+}$, is injective.
\end{defn}
\hspace{0pt}\indent
	As specified by the definition above, an adaptive code of order $n$ associates variable-length
	codewords to symbols being encoded depending on the previous $n$ symbols in the input data string.
	Let us take an example in order to better understand this adaptive mechanism.
\begin{exmp}
	This example makes use of an adaptive code of order two.
	Consider $\Sigma=\{\textup{a,b,c}\}$ and $\Delta=\{\textup{0,1}\}$ two alphabets, and
	$\ac{2}$ a function constructed by the following table.
	One can easily verify that $\overline{c}$ is injective, and according to Definition 1, $c$ is
	an adaptive code of order two.
\newline
\begin{table}
\begin{center}
Table 1. An adaptive code of order two.
\begin{tabular}{|c|c|c|c|c|c|c|c|c|c|c|c|c|c|}
\hline
$\Sigma\backslash\Sigma^{\leq{2}}$	& a  & b	& c	& aa	& ab & ac & ba & bb & bc & ca & cb & cc & $\lambda$\\ \hline
a 					& 01 & 10 	& 10 	& 00 	& 11 & 10 & 01 & 10 & 11 & 11 & 11 & 00 & 00	   \\ \hline
b					& 10 & 00	& 11	& 11	& 01 & 00 & 00 & 11 & 01 & 10 & 00 & 10 & 11       \\ \hline
c					& 11 & 01	& 01	& 10	& 00 & 11 & 11 & 00 & 00 & 00 & 10 & 11 & 10	   \\ \hline
\end{tabular}
\end{center}	
\end{table}
\newline
	Let $x=\textup{abacca}\in\Sigma^{+}$ be an input data string. 
	Using the definition above, we encode $x$ by
	$\overline{c}(x)=
	c(\textup{a,$\lambda$})c(\textup{b,a})c(\textup{a,ab})c(\textup{c,ba})c(\textup{c,ac})c(\textup{a,cc})=
	001011111100$.
\end{exmp}
\hspace{0pt}\indent
	Let $\ac{n}$ be an adaptive code of order $n$, $n\geq{1}$. We denote by
	$C_{c, \sigma_{1}\sigma_{2}\ldots\sigma_{h}}$ the set
	$\{c(\sigma,\sigma_{1}\sigma_{2}\ldots\sigma_{h}) \mid \sigma\in\Sigma\}$,
	for all $\sigma_{1}\sigma_{2}\ldots\sigma_{h}\in\Sigma^{\leq{n}}-\{\lambda\}$,
	and by $C_{c, \lambda}$ the set $\{c(\sigma,\lambda) \mid \sigma\in\Sigma\}$.
	We write $C_{\sigma_{1}\sigma_{2}\ldots\sigma_{h}}$ instead of
	$C_{c, \sigma_{1}\sigma_{2}\ldots\sigma_{h}}$,
	and $C_{\lambda}$ instead of $C_{c, \lambda}$
	whenever there is no confusion.
	Let us denote by $AC(\Sigma,\Delta,n)$ the set
	$\{\ac{n} \mid $ c is an adaptive code of order $n$ $\}$.
	The proof of the following theorem can be found in \cite{t1}.
\begin{thm}
	Let $\Sigma$ and $\Delta$ be two alphabets, and $\ac{n}$ a function, $n\geq{1}$.
	If $C_{u}$ is prefix code, for all $u\in\Sigma^{\leq{n}}$, then $c\in{AC(\Sigma,\Delta,n)}$.
\end{thm}

\section{Translating the EAH Encoder into Automata Theory}
	In this section, we focus on translating the EAH data compression algorithm \cite{t3} into automata theory.
	Before translating the algorithm, some new definitions are needed.
	For further details on formal languages, the reader is reffered to \cite{rs1}.
\begin{defn}
	Let $\Sigma$ be an alphabet such that $\star\notin\Sigma$,
	and $w=u_{1}u_{2}\ldots{u_{h}}\in\Sigma^{+}$, $u_{i}\in\Sigma$,
	$\forall{i}$. The \textup{adaptive automaton of order $n$} associated to $w$
	is the nondeterministic finite automaton
	$A_{n}(w)=(S(w,n),T(w,n),\delta_{w,n},s_{0}(w,n),F(w,n))$, where
	$S(w,n),T(w,n),\delta_{w,n},s_{0}(w,n),$ and $F(w,n)$ are defined by:
\begin{itemize}
\item[-]$S(w,n)=\{u_{j} \mid n+1\leq{j}\leq{h}\}\cup\{u_{j}u_{j+1}\ldots{u_{j+n-1}} \mid
	1\leq{j}\leq{h-n}\}\cup\{\star\};$
\item[-]$T(w,n)=\{(a,code) \mid \exists$ $j$, $1\leq{j}\leq{h-n},$  \textup{such that}
	$a=|\{k \mid 1\leq{k}\leq{h-n},$ 
	$u_{k}u_{k+1}\ldots{u_{k+n}}=u_{j}u_{j+1}\ldots{u_{j+n}}\}|,$
	$code\in\{0,1\}^{+}$ \textup{is the codeword associated to} $u_{j+n}$,
	\textup{when the previous} $n$ \textup{symbols are} $u_{j}\ldots{u_{j+n-1}}\}\cup$
	\newline
	$\{(a,\lambda) \mid \exists$ $j,$ $n\leq{j}\leq{h}$ \textup{such that}
	$a=|\{k \mid u_{k-n+1}\ldots{u_{k}}=u_{j-n+1}\ldots{u_{j}}\}|\};$
\item[-]$\delta_{w,n}:S(w,n)\times{T(w,n)}\rightarrow{\mathcal{P}(S(w,n))}$ is given by:
\item[ ]$\delta_{w,n}(s,(a,c))=
	\left\{
	\begin{array}{ll}
	\{\star\}			& \textrm{\textup{if} $s=\star$}		\\
	Z_{w}^{n}(s,a,c)		& \textrm{\textup{if} $s\neq\star$ \textup{and}
						$Z_{w}^{n}(s,a,c)\neq\emptyset$}		\\
	\{\star\}			& \textrm{\textup{if} $s\neq\star$ \textup{and}
						$Z_{w}^{n}(s,a,c)=\emptyset$}		\\
	\end{array}
	\right.$
\item[-]$s_{0}(w,n)=u_{1}u_{2}\ldots{u_{n}};$
\item[-]$F(w,n)=\{u_{h-n+1}u_{h-n+2}\ldots{u_{h}}\};$
\end{itemize}
	and $Z_{w}^{n}(s,a,c)$ is given by:
\newline\newline
	$Z_{w}^{n}(s,a,c)=
	\left\{
	\begin{array}{ll}
	\{p \mid \exists j:$ $u_{j}u_{j+1}=sp,$
	$p$ \textup{is}
	$
					& \textrm{}								\\
	$
	\textup{encoded by} $c,$ $a=|\{k \mid u_{k}u_{k+1}=sp\}|\}$
	$
					& \textrm{\textup{if} $n=1,$ $|s|=1.$}			\\
	\{p \mid \exists j:$ $u_{j}\ldots{u_{j+n}}=sp,$ $|p|=1,$
	$p$ \textup{is}
	$
					& \textrm{}								\\
	$
	\textup{encoded by} $c,$ $a=|\{k \mid u_{k}\ldots{u_{k+n}}=sp\}|\}$
	$
					& \textrm{\textup{if} $n\geq{2},$ $|s|\geq{2}.$}	\\
	\{p \mid \exists j:$ $u_{j-n+1}\ldots{u_{j}}=p,$ $u_{j}=s,$ 
	$c=\lambda,$
	$
					& \textrm{}								\\
	$
	$a=|\{k \mid u_{k-n+1}\ldots{u_{k}}=p,$ $u_{k}=s\}|\}$
	$
					& \textrm{\textup{if} $n\geq{2},$ $|s|=1.$}		\\
	\end{array}
	\right.$
\end{defn}
\begin{exmp}
	Let $\Sigma=\{\textup{a,b,c,d}\}$ be an alphabet, and $w=\textup{abdbacdba}\in\Sigma^{+}$.
	It is easy to verify
	that the adaptive automaton of order one associated to $w$ is constructed as below
	(the algorithm which associates the codewords is not important in this example).
\newline
\setlength{\unitlength}{1pt}
\begin{picture}(395,170)
	\put(20,10){\circle{20}}
	\put(20,130){\circle{20}}
	\put(375,10){\circle{20}}
	\put(375,130){\circle{20}}
	\put(20,130){\circle{23}}
	\put(197.5,50){\circle{20}}	

	\put(20,157.5){\vector(0,-1){15}}

	\put(17.5,127.5){\textup{a}}
	\put(17.5,7.2){\textup{c}}
	\put(372.5,7.3){\textup{d}}
	\put(372.5,127.1){\textup{b}}
	\put(194.3,47){\large{$\star$}}
	
	\put(20,118){\vector(0,-1){98}}
	\put(30,10){\vector(1,0){335}}
	\put(31.5,126){\vector(1,0){334}}
	\put(365.5,134){\vector(-1,0){334}}
	\put(379,19.5){\vector(0,1){101}}
	\put(371,120.5){\vector(0,-1){101}}

	\put(28,121){\line(0,-1){67}}
	\put(28,54){\vector(1,0){160}}
	\put(28,15.5){\line(0,1){29.5}}
	\put(28,45){\vector(1,0){160}}

	\put(367,123){\line(-1,0){30}}
	\put(367,16){\line(-1,0){30}}
	\put(337,123){\line(0,-1){69}}
	\put(337,16){\line(0,1){29}}
	\put(337,54){\vector(-1,0){130}}
	\put(337,45){\vector(-1,0){130}}

	\put(192,58){\line(0,1){15}}
	\put(203,73){\vector(0,-1){15}}
	\put(192,73){\line(-1,0){50}}
	\put(203,73){\line(1,0){50}}
	\put(142,73){\line(0,1){6}}
	\put(253,73){\line(0,1){6}}
	\put(142,79){\line(1,0){111}}
	
	\put(8,60){\rotatebox{90}{\scriptsize{$(1,1)$}}}
	\put(381,60){\rotatebox{90}{\scriptsize{$(2,0)$}}}
	\put(188,116){\scriptsize{$(1,0)$}}
	\put(188,139){\scriptsize{$(2,0)$}}
	\put(359,60){\rotatebox{90}{\scriptsize{$(1,1)$}}}
	\put(187,14){\scriptsize{$(1,0)$}}
	\put(70,59){\scriptsize{$(2,0)$}}
	\put(70,36){\scriptsize{$(1,1),(2,0)$}}
	\put(270,59){\scriptsize{$(1,0)$}}
	\put(270,36){\scriptsize{$(1,0),(1,1)$}}
	\put(165,84){\scriptsize{$(1,0),(1,1),(2,0)$}}
\end{picture}
\begin{center}
	\small{\textup{Fig. 1.} $A_{1}(\textup{abdbacdba})$}
\vspace{5pt}
\end{center}
\end{exmp}
	Let $U=(u_{1},u_{2},\ldots,u_{k})$ be a $k$-tuple.
	We denote by $(u_{1},u_{2},\ldots,u_{k}).i$ the $i$-th component of $U$,
	that is, $u_{i}=(u_{1},u_{2},\ldots,u_{k}).i$, for all $i$, $1\leq{i}\leq{k}$.
	The $0$-tuple is denoted by $()$. The length of the tuple $U$ is denoted by $Len(U)$.
	If $V=(v_{1},v_{2},\ldots,v_{b})$,
	$M=(m_{1},m_{2},\ldots,m_{r},U)$, $N=(n_{1},n_{2},\ldots,n_{s},V)$,
	$P=(p_{1},\ldots,p_{i-1},p_{i},p_{i+1},\ldots,p_{t})$ are tuples and q is an element or a tuple,
	then we define
	$P\vartriangleleft{q}$, $P\vartriangleright{i}$, $U\vartriangle{V}$, and $M\lozenge{N}$ by:
\begin{itemize}
\item $P\vartriangleleft{q}=(p_{1},\ldots,p_{t},q)$
\item $P\vartriangleright{i}=(p_{1},\ldots,p_{i-1},p_{i+1},\ldots,p_{t})$
\item	$U\vartriangle{V}=(u_{1},u_{2},\ldots,u_{k},v_{1},v_{2},\ldots,v_{b})$
\item $M\lozenge{N}=(m_{1}+n_{1},m_{2}+1,\ldots,m_{r}+1,n_{2}+1,\ldots,n_{s}+1,U\vartriangle{V})$
\end{itemize}
	where $m_{i}$, $n_{j}$ are integers, $1\leq{i}\leq{r}$ and $1\leq{j}\leq{s}$.
	If $(f_{1},\ldots,f_{k})$ is a tuple of integers, let us denote by Huffman$(f_{1},\ldots,f_{k})$
	the Huffman algorithm, which returns a tuple $((c_{1},l_{1}),\ldots,(c_{k},l_{k}))$,
	where $c_{i}$ is the codeword associated to the symbol with the frequency $f_{i}$,
	and $l_{i}$ is the length of $c_{i}$, $1\leq{i}\leq{k}$.
\begin{alg}
\textup{
\hspace{0pt}
\newline\newline
$\mathbf{Huffman}$(tuple $\mathcal{F}=(f_{1},f_{2},\ldots,f_{k})$, where $k\geq{1}$)
\newline
$\texttt{1.}$ $\mathcal{L}:=((f_{1},0,(1)),(f_{2},0,(2)),\ldots,(f_{k},0,(k)));$
\newline
$\texttt{2.}$ Let $\mathcal{V}=(\lambda,\lambda,\ldots,\lambda)$ be a $k$-tuple;
\newline
$\texttt{3.}$ \textbf{if} $k=1$ \textbf{then} $\mathcal{V}.1:=0;$
\newline
$\texttt{4.}$ \textbf{while} $Len(\mathcal{L})>1$ \textbf{do}
\newline\vspace{0pt}
\hspace{12pt}\textbf{begin}
\newline
$\texttt{4.1.}$
	Let $i<j$ be such that $1\leq{i,j}\leq{Len(\mathcal{L})}$ and $\mathcal{L}.i.1$, $\mathcal{L}.j.1$ are
\newline\vspace{0pt}
\hspace{25pt}the smallest elements of the set $\{\mathcal{L}.q.1 \mid 1\leq{q}\leq{Len(\mathcal{L})}\};$
\newline
$\texttt{4.2.}$
$First:=\{\mathcal{L}.i.Len(\mathcal{L}.i).r\mid 1\leq{r}\leq{Len(\mathcal{L}.i.Len(\mathcal{L}.i))}\};$
\newline
$\texttt{4.3.}$
$Second:=\{\mathcal{L}.j.Len(\mathcal{L}.j).r\mid 1\leq{r}\leq{Len(\mathcal{L}.j.Len(\mathcal{L}.j))}\};$
\newline
$\texttt{4.4.}$ \textbf{for} each $x\in{First}$ \textbf{do} $\mathcal{V}.x:=0\cdot{\mathcal{V}.x};$
\newline
$\texttt{4.5.}$ \textbf{for} each $x\in{Second}$ \textbf{do} $\mathcal{V}.x:=1\cdot{\mathcal{V}.x};$
\newline
$\texttt{4.6.}$ 	$\mathcal{U}:=\mathcal{L}.i$ $\lozenge$ $\mathcal{L}.j;$
		$\mathcal{L}:=\mathcal{L}\vartriangleright{j};$
		$\mathcal{L}:=\mathcal{L}\vartriangleright{i};$
		$\mathcal{L}:=\mathcal{L}\vartriangleleft{\mathcal{U}};$
\newline\vspace{0pt}
\hspace{12pt}\textbf{end}
\newline
$\texttt{5.}$ \textbf{return} $((\mathcal{V}.1,|\mathcal{V}.1|),\ldots,(\mathcal{V}.k,|\mathcal{V}.k|));$
}
\end{alg}
	Let $\Sigma=\{\sigma_{0},\sigma_{1},\ldots,\sigma_{m-1}\}$ be an alphabet, $1\leq{m}\leq{256}$.
	Let us recall the idea of our algorithm, denoted by $\textup{EAH}n$.
	For example, let $w=w_{1}w_{2}\ldots{w_{h}}$ be a string
	over $\Sigma$. We encode $w$ by a $5$-tuple $U=(A,B,C,D,E)$, where $A,B,C,D,E$ are bitstrings
	constructed as below.

	$\mathbf{EAH}n(w).1=A.$\hspace{18pt}
	Let $Index:\Sigma\rightarrow\{0,1,\ldots,m-1\}$ be a function which gets as input
	a symbol $\sigma\in\Sigma$ and outputs an index $i$ such that $\sigma=\sigma_{i}$.
	If $h\geq{n}$, then $A=Base10Base2(Index(w_{1}))\ldots Base10Base2(Index(w_{n}))$, that is,
	$A$ is the conversion of the sequence $Index(w_{1}),\ldots,Index(w_{n})$ from base 10
	to base 2, and $|A|=n*\lceil\log_{2}m\rceil$. Otherwise, if $h<n$, then we consider
	$A=Base10Base2(Index(w_{1}))\ldots Base10Base2(Index(w_{h}))$, that is,
	$A$ is the conversion of $Index(w_{1}),\ldots,Index(w_{h})$ from base 10
	to base 2, and $|A|=h*\lceil\log_{2}m\rceil$.

	$\mathbf{EAH}n(w).2=B.$\hspace{18pt}
	$B=B_{0}B_{1}\ldots{B_{m^{n}-1}}$, where $B_{j}$ is defined by:
\begin{displaymath}
	B_{j}=
	\left\{
		\begin{array}{ll}
		1	& \textrm{if $\exists$ $i\in\{0,\ldots,m-1\}$ and $\exists$ $k\in\{1,\ldots,h\}$ 
			such that}\\
			& \textrm{$Index^{-1}(10\_m(j).1))\ldots{Index^{-1}(10\_m(j).n)\cdot\sigma_{i}}$
			= $w_{k}\ldots{w_{k+n}}$}	\\
		0	& \textrm{otherwise}
		\end{array}
	\right.
\end{displaymath}
	for all $j$, $0\leq{j}\leq{m^{n}-1}$, and $10\_m(j)$ is a tuple of length $n$ denoting
	the conversion of $j$ from base 10 to base $m$, such that $Len(10\_m(j))=n$ and $10\_m(j).i$ is
	the $i$-th digit (from left to right) of this conversion.

	$\mathbf{EAH}n(w).3=C.$\hspace{18pt}
	$C=C_{0}C_{1}\ldots{C_{m-1}}$, where $C_{i}=C_{i}^{0}C_{i}^{1}\ldots{C_{i}^{m^{n}-1}}$,
	$0\leq{i}\leq{m-1}$, and for all $j$, $0\leq{j}\leq{m^{n}-1},$ $C_{i}^{j}\in\{0,1\}^{*}$ is defined by:
\begin{displaymath}
	C_{i}^{j}=
	\left\{
		\begin{array}{ll}
		1		& \textrm{if $B_{j}=1$ and $\exists$ $k\in\{1,2,\ldots,h\}$ such that}\\
				& \textrm{$Index^{-1}(10\_m(j).1))\ldots{Index^{-1}(10\_m(j).n)\cdot\sigma_{i}}$
					= $w_{k}\ldots{w_{k+n}}.$}	\\
		0		& \textrm{if $B_{j}=1$ and $\nexists$ $k\in\{1,2,\ldots,h\}$ such that}\\
				& \textrm{$Index^{-1}(10\_m(j).1))\ldots{Index^{-1}(10\_m(j).n)\cdot\sigma_{i}}$
					= $w_{k}\ldots{w_{k+n}}.$}	\\
		\lambda	& \textrm{if $B_{j}=0.$}
		\end{array}
	\right.
\end{displaymath}
	$\mathbf{EAH}n(w).4=D.$\hspace{18pt}
	$D=D_{0}D_{1}\ldots{D_{m-1}}$, where $D_{i}=D_{i}^{0}D_{i}^{1}\ldots{D_{i}^{m^{n}-1}}$,
	$0\leq{i}\leq{m-1}$, and $D_{i}^{j}\in\{0,1\}^{*}$ is defined by:
\begin{displaymath}
	D_{i}^{j}=
	\left\{
		\begin{array}{ll}
		MBase10Base2(Freq(\sigma_{i},j))			& \textrm{if $C_{i}^{j}=1$}		\\
		\lambda							& \textrm{if $C_{i}^{j}\neq{1}$}
		\end{array}
	\right.
\end{displaymath}
	where
	$Freq(\sigma_{i},j)=|\{k\mid 
		w_{k}\ldots{w_{k+n}}=\sigma_{10\_m(j).1}\ldots{\sigma_{10\_m(j).n}}\sigma_{i}\}|$.

	Let us denote by $Marked$ the set $\{(i,j)\mid C_{i}^{j}=1\}$.
	The greatest element of the set
	$\{|Base10Base2(Freq(\sigma_{i},j))|\mid (i,j)\in{Marked}\}$ is denoted by $Max$.
	Then, $MBase10Base2(Freq(\sigma_{i},j))$ is defined by:
\begin{displaymath}
	MBase10Base2(Freq(\sigma_{i},j)) = \underbrace{00\ldots{0}}_{t(i,j)}Base10Base2(Freq(\sigma_{i},j))
\end{displaymath}
	where $t(i,j)=Max-|Base10Base2(Freq(\sigma_{i},j))|$.

	$\mathbf{EAH}n(w).5=E.$\hspace{18pt}
	$E$ denotes the compression of $w$ using $A,B,C,D$, the adaptive automaton of order $n$
	associated to $w$, and the Huffman algorithm.
\begin{alg}
\textup{
\hspace{0pt}
\newline\newline
\textbf{EAH}$n$(string $w=w_{1}w_{2}\ldots{w_{h}}\in\Sigma^{+}$, such that $w_{i}\in\Sigma$, $1\leq{i}\leq{h}$,
			$h>n$)
\newline
$\texttt{ 1.}$
	$A:=\lambda;B:=\lambda;C=\lambda;D:=\lambda;E:=\lambda;A_{n}(w).1:=\{\star\};A_{n}(w).2:=\emptyset;$
\newline
$\texttt{ 2.}$ 	$A_{n}(w).4:=w_{1}\ldots{w_{n}};A_{n}(w).5:=\{w_{h-n+1}\ldots{w_{h}}\};M_{1}:=();M_{2}:=();$
\newline
$\texttt{ 3.}$ 	\textbf{for} $j=n+1$ \textbf{to} $h$ \textbf{do}
			$A_{n}(w).1:=A_{n}(w).1\cup\{w_{j}\};$
\newline
$\texttt{ 4.}$	\textbf{for} $j=1$ \textbf{to} $h-n$ \textbf{do}
			$A_{n}(w).1:=A_{n}(w).1\cup\{w_{j}w_{j+1}\ldots{w_{j+n-1}}\};$
\newline
$\texttt{ 5.}$ 	\textbf{for} $j=n+1$ \textbf{to} $h$ \textbf{do}
\newline
$\texttt{ 6.}$
	\hspace{10pt}\textbf{if} $\exists$ $i$ such that $M_{1}.i.1=w_{j-n}\ldots{w_{j-1}}$ and $M_{1}.i.2=w_{j}$
\newline
$\texttt{ 7.}$
	\hspace{20pt}\textbf{then} $M_{1}.i.3:=M_{1}.i.3+1;$
\newline
$\texttt{ 8.}$
	\hspace{20pt}\textbf{else} $M_{1}:=M_{1}\vartriangleleft(w_{j-n}\ldots{w_{j-1}},w_{j},0,\lambda);$
\newline
$\texttt{ 9.}$ $x:=|\{i \mid M_{1}.i.3=0\}|;$
\newline
$\texttt{10.}$ \textbf{while} $x>0$ \textbf{do}
\newline
\vspace{0pt}\hspace{18pt} \textbf{begin}
\newline
$\texttt{10.1.}$ $pos:=min\{i \mid M_{1}.i.3=0\};$ $\mathcal{U}:=(M_{1}.pos);$ $\mathcal{V}:=(pos);$
\newline
$\texttt{10.2.}$ \textbf{for} $j=pos+1$ \textbf{to} $Len(M_{1})$ \textbf{do}
\newline
$\texttt{10.3.}$
	\hspace{10pt}\textbf{if} $M_{1}.j.1=M_{1}.pos.1$ \textbf{then}
\newline
\vspace{0pt}\hspace{41pt} \textbf{begin}
\newline
$\texttt{10.3.1.}$
	\hspace{11pt}$\mathcal{U}:=\mathcal{U}\vartriangleleft{M_{1}.j};$
\newline
$\texttt{10.3.2.}$
	\hspace{11pt}$\mathcal{V}:=\mathcal{V}\vartriangleleft{j};$
\newline
\vspace{0pt}\hspace{41pt} \textbf{end}
\newline
$\texttt{10.4.}$ \textbf{for} $j=1$ \textbf{to} $Len(\mathcal{U})$ \textbf{do}
	$\mathcal{U}.j.4:=$Huffman$((\mathcal{U}.1.3,\ldots{\mathcal{U}.Len(\mathcal{U}).3})).j.1;$
\newline
$\texttt{10.5.}$ \textbf{for} $j=1$ \textbf{to} $Len(\mathcal{V})$ \textbf{do}
	$M_{1}.(\mathcal{V}.j).3:=\mathcal{U}.j.3;$
\newline
$\texttt{10.6.}$ $x:=|\{i \mid M_{1}.i.3=0\}|;$
\newline
\vspace{0pt}\hspace{18pt} \textbf{end}
\newline
$\texttt{11.}$ \textbf{for} $j=n$ \textbf{to} $h$ \textbf{do}
\newline
$\texttt{12.}$ \textbf{if} $\exists$ $i$ such that $M_{2}.i.1=w_{j}$ and $M_{2}.i.2=w_{j-n+1}\ldots{w_{j}}$
\newline
$\texttt{13.}$
	\hspace{10pt}\textbf{then} $M_{2}.i.3:=M_{2}.i.3+1;$
\newline
$\texttt{14.}$
	\hspace{10pt}\textbf{else} $M_{2}:=M_{2}\vartriangleleft(w_{j},w_{j-n+1}\ldots{w_{j}},0,\lambda);$
\newline
$\texttt{15.}$ \textbf{for} $j=1$ \textbf{to} $Len(M_{1})$ \textbf{do}
	$A_{n}(w).2:=A_{n}(w).2\cup\{(M_{1}.j.3,M_{1}.j.4)\};$
\newline
$\texttt{16.}$ \textbf{for} $j=1$ \textbf{to} $Len(M_{2})$ \textbf{do}
	$A_{n}(w).2:=A_{n}(w).2\cup\{(M_{2}.j.3,M_{2}.j.4)\};$
\newline
$\texttt{17.}$
	\textbf{for} each $t\in{A_{n}(w).2}$ \textbf{do}
	$A_{n}(w).3(\star,t):=\{\star\};$
\newline
$\texttt{18.}$ \textbf{for} each $(s,t)\in{(A_{n}(w).1-\{\star\})\times{A_{n}(w).2}}$ \textbf{do}
	$A_{n}(w).3(s,t):=\emptyset;$
\newline
$\texttt{19.}$ \textbf{for} each $(s,t)\in{(A_{n}(w).1-\{\star\})\times{A_{n}(w).2}}$ \textbf{do}
\newline
$\texttt{20.}$
	\hspace{10pt}\textbf{for} $j=1$ \textbf{to} $Len(M_{1}\vartriangle{M_{2}})$ \textbf{do}
\newline
$\texttt{21.}$
	\hspace{20pt}\textbf{if} $(M_{1}\vartriangle{M_{2}}).j.1=s$
	and
	$t=((M_{1}\vartriangle{M_{2}}).j.3,(M_{1}\vartriangle{M_{2}}).j.4)$
\newline
$\texttt{22.}$
	\hspace{30pt}\textbf{then} $A_{n}(w).3(s,t):=A_{n}(w).3(s,t)\cup\{(M_{1}\vartriangle{M_{2}}).j.2\};$
\newline
$\texttt{23.}$ \textbf{for} each $(s,t)\in{(A_{n}(w).1-\{\star\})\times{A_{n}(w).2}}$ \textbf{do}
\newline
$\texttt{24.}$
	\hspace{10pt}\textbf{if} $A_{n}(w).3(s,t)=\emptyset$
\newline
$\texttt{25.}$
	\hspace{20pt}\textbf{then} $A_{n}(w).3(s,t)=\{\star\};$
\newline
$\texttt{26.}$ $A:=Base10Base2(Index(w_{1}))\ldots{Base10Base2(Index(w_{n}))};$
\newline
$\texttt{27.}$ \textbf{for} $j=0$ \textbf{to} $m^{n}-1$ \textbf{do}
\newline
$\texttt{28.}$
	\hspace{10pt}\textbf{if} $\exists$ $t\in{A_{n}(w).2}$ such that
	$\delta(\sigma_{10\_m(j).1}\ldots{\sigma_{10\_m(j).n}},t)-\{\star\}\neq\emptyset$
\newline
$\texttt{29.}$
	\hspace{20pt}\textbf{then} $B:=B\cdot{1};$
\newline
$\texttt{30.}$
	\hspace{20pt}\textbf{else} $B:=B\cdot{0};$
\newline
$\texttt{31.}$ \textbf{for} $i=0$ \textbf{to} $m-1$ \textbf{do}
\newline
$\texttt{32.}$
	\hspace{10pt}\textbf{for} $j=0$ \textbf{to} $m^{n}-1$ \textbf{do}
\newline
$\texttt{33.}$
	\hspace{20pt}\textbf{if} $B_{j}=1$ \textbf{then}
\newline
$\texttt{34.}$
	\hspace{30pt}\textbf{if}
		$\exists$ $t$ such that
		$\delta(\sigma_{10\_m(j).1}\ldots{\sigma_{10\_m(j).n}},t)=\sigma_{i}$
	\textbf{then}
\newline
\vspace{0pt}\hspace{48pt} \textbf{begin}
\newline
$\texttt{34.1.}$
	\hspace{26pt} $C:=C\cdot{1};$
\newline
$\texttt{34.2.}$
	\hspace{26pt} $D:=D\cdot{MBase10Base2(t.1)};$
\newline
\vspace{0pt}\hspace{48pt} \textbf{end}
\newline
$\texttt{35.}$
	\hspace{30pt}\textbf{else} $C:=C\cdot{0};$
\newline
$\texttt{36.}$ \textbf{for} $i=n+1$ \textbf{to} $h$ \textbf{do}
\newline
\vspace{0pt}\hspace{18pt} \textbf{begin}
\newline
$\texttt{36.1.}$ Let $t\in{A_{n}(w).2}$ be such that
	$w_{i}\in\delta(w_{i-n}\ldots{w_{i-1}},t);$
\newline
$\texttt{36.2.}$ $E:=E\cdot{t.2};$
\newline
\vspace{0pt}\hspace{18pt} \textbf{end}
\newline
$\texttt{37.}$ \textbf{return} $(A,B,C,D,E);$
}
\end{alg}
	Let $u$ be a string. In the remainder of this paper, we denote by
	$\textup{LZ}(u)$ the encoding of $u$ using the Lempel-Ziv data compression algorithm \cite{zl1},
	and by $\textup{H}(u)$ the encoding of $u$ using the Huffman algorithm. Also, let us consider the
	following notations:
\begin{itemize}
\item $\textup{LH}(u)=|\textup{H}(u)|$;
\item $\textup{LLZ}(u)=|\textup{LZ}(u)|;$
\item $\textup{LEAH}n(u)=\sum_{i=1}^{5}|\textup{EAH}n(u).i|.$
\end{itemize}
\begin{exmp}
	Let $\Sigma=\{\textup{a,b,c,d,e}\}$ be an alphabet, and $w$ a string of length $200$ over $\Sigma$, given
	as below (between brackets).

\textup{
	$w$\hspace{0.3pt}=
	[\texttt{abedcababedccabedcedcababedcedcccabedcabedcedccababedcabedc}
\newline
	\texttt{cccedccedccedcababedcabedcedccedcababedcabedccabedcababedcedcccc}
\newline
	\texttt{cedcabedcabedccccedcccabedcccedccabedccccabedccababedcabedcedcca}
\newline
	\texttt{bedcababedced}]}

	Applying $\textup{EAH}1$ to the input string $w$, we get the following adaptive automaton of
	order one.
\newline
\setlength{\unitlength}{1pt}
\begin{picture}(395,145)
	\put(381,140){\vector(0,-1){10}}

	\put(14,120){\circle{20}}
	\put(381,120){\circle{20}}
	\put(11.3,117){\textup{b}}
	\put(378.5,117.5){\textup{a}}
	\put(23.7,117){\vector(1,0){347.3}}
	\put(370.7,123){\vector(-1,0){347.3}}
	
	\put(64,90){\circle{20}}
	\put(331,90){\circle{20}}
	\put(61.3,87){\textup{e}}
	\put(328.4,87){\textup{c}}
	\put(321,90){\vector(-1,0){247}}

	\put(197.5,60){\circle{20}}
	\put(197.5,10){\circle{20}}
	\put(194.8,57){\textup{d}}
	\put(194.8,7){\large{$\star$}}

	\put(14,110){\line(0,-1){106}}
	\put(14,4){\vector(1,0){175}}
	\put(381,110){\line(0,-1){106}}
	\put(381,4){\vector(-1,0){175}}

	\put(22,113){\line(1,0){42}}
	\put(64,113){\vector(0,-1){13}}

	\put(331,100){\line(0,1){13}}
	\put(331,113){\vector(1,0){42}}

	\put(326,99){\line(0,1){14}}
	\put(326,113){\line(-1,0){20}}
	\put(306,113){\line(0,-1){18}}
	\put(306,95){\vector(1,0){16}}

	\put(73,85){\line(1,0){120}}
	\put(193,85){\vector(0,-1){16}}

	\put(202,69){\line(0,1){16}}
	\put(202,85){\vector(1,0){120}}

	\put(64,80){\line(0,-1){70}}
	\put(64,10){\vector(1,0){123.5}}

	\put(331,80){\line(0,-1){70}}
	\put(331,10){\vector(-1,0){123.5}}
	
	\put(188,55){\line(-1,0){100}}
	\put(88,55){\line(0,-1){8}}
	\put(88,47){\line(1,0){80}}
	\put(168,47){\line(0,-1){31}}
	\put(168,16){\vector(1,0){21}}

	\put(197.5,20){\line(0,1){20}}
	\put(197.5,40){\line(1,0){19.5}}
	\put(217,40){\line(0,1){15}}
	\put(217,55){\line(1,0){90}}
	\put(307,55){\line(0,-1){8}}
	\put(307,47){\line(-1,0){80}}
	\put(227,47){\line(0,-1){31}}
	\put(227,16){\vector(-1,0){21}}

	\put(197.5,60){\circle{17}}

	\put(187,126){\scriptsize{$(31,0)$}}
	\put(189,110){\scriptsize{$(8,0)$}}
	\put(30,106){\scriptsize{$(23,1)$}}
	\put(340,106){\scriptsize{$(22,10)$}}
	\put(297,92){\rotatebox{90}{\scriptsize{$(28,0)$}}}
	\put(185,93){\scriptsize{$(14,11)$}}
	\put(87,78){\scriptsize{$(37,0)$}}
	\put(285,78){\scriptsize{$(36,0)$}}
	
	\put(4,5){\rotatebox{90}{\scriptsize{$(31,0),(37,0),(14,11)$}}}
	\put(17,5){\rotatebox{90}{\scriptsize{$(36,0),(28,0),(22,10)$}}}

	\put(371,5){\rotatebox{90}{\scriptsize{$(23,1),(37,0),(14,11)$}}}
	\put(382,5){\rotatebox{90}{\scriptsize{$(8,0),(36,0),(28,0),(22,10)$}}}

	\put(54,20){\rotatebox{90}{\scriptsize{$(31,0),(8,0)$}}}
	\put(65,20){\rotatebox{90}{\scriptsize{$(23,1),(14,11)$}}}
	\put(75,14){\scriptsize{$(36,0),(28,0),(22,10)$}}

	\put(87,59){\scriptsize{$(23,1),(37,0),(14,11)$}}
	\put(87,40){\scriptsize{$(31,0),(8,0),(28,0)$}}
	\put(169,46){\rotatebox{270}{\scriptsize{$(22,10)$}}}

	\put(220,59){\scriptsize{$(23,1),(37,0),(14,11)$}}
	\put(199,20){\rotatebox{90}{\scriptsize{$(8,0)$}}}
	\put(218,17){\rotatebox{90}{\scriptsize{$(31,0)$}}}
	\put(229,40){\scriptsize{$(36,0),(28,0),(22,10)$}}

	\put(332,20){\rotatebox{90}{\scriptsize{$(23,1),(37,0)$}}}
	\put(245,14){\scriptsize{$(31,0),(36,0),(8,0)$}}

\end{picture}
\begin{center}
	\small{\textup{Fig. 2.} $A_{1}(w)$}
\vspace{5pt}
\end{center}
	One can easily verify that we obtain:
\begin{center}
	$
	\textup{LH}(w)=462\hspace{50pt}
	\textup{LEAH}1(w)=310\hspace{50pt}
	\textup{LLZ}(w)=388
	$
\end{center}
	which shows that in this case, $\textup{EAH}1$ substantially outperforms the well-known
	Lempel-Ziv universal data compression algorithm \cite{zl1}.
\end{exmp}

\section{Concluding Remarks}
	The EAH data compression algorithm has been proposed in \cite{t3}, where we have behaviorally
	shown that for a large class of input data strings, this algorithm substantially outperforms
	the Lempel-Ziv data compression algorithm. In this paper, we translated
	the EAH algorithm into automata theory.
	Further work on adaptive codes will be focused on finding new improvements for the
	EAH algorithm, as well as other algorithms for data compression based on adaptive codes.






\end{document}